\documentclass[fleqn,11pt]{wlscirep}
\usepackage{graphicx,amssymb,amsmath,amsthm,cite,cases}
\usepackage{adjustbox}
\usepackage{algorithmic,setspace}
\usepackage{color,multirow}
\usepackage{mathrsfs}
\usepackage[ruled]{algorithm}

\def\til{\tilde}
\def\ga{\gamma}

\def\vst{\mathbf{\til{s}}}
\def\vstr{\mathbf{\til{s}}^{\rm{r}}}

\def\ellt{\til{\ell}}
\def\vct{\mathbf{\til{c}}}
\def\vdelt{\boldsymbol{\til{\delta}}}
\def\vs{\mathbf{s}}
\def\vt{\mathbf{t}}
\def\vga{\boldsymbol{\gamma}}
\def\vgat{\boldsymbol{\til{\gamma}}}
\def\vll{\boldsymbol{\ell}}
\def\vllt{\boldsymbol{\til{\ell}}}
\def\be{\begin{equation}}
\def\ee{\end{equation}}
\def\ben{\begin{eqnarray}}
\def\een{\end{eqnarray}}

\def\R{\mathbb{R}}

\def\CRAH{\mathrm{\CR^{\mathrm{Huff}}_{\mathrm{a}}}}
\def\CRBH{\mathrm{\CR^{\mathrm{Huff}}_{\mathrm{b}}}}

\def\CRB{\mathrm{CR_{\mathrm{b}}}}
\def\CR{\mathrm{CR}}
\def\CRA{\mathrm{CR_{\mathrm{a}}}}
\def\CRRL{\mathrm{CR_{\mathrm{RL}}}}
\def\CRRLH{\mathrm{\CR^{\mathrm{Huff}}_{\mathrm{RL}}}}

\def\vf{\mathbf{f}}

\def\vc{\mathbf{c}}
\def\vcq{\mathbf{c}^\Delta}
\def\vcqt{\til{\mathbf{c}}^\Delta}

\def\vw{\mathbf{w}}
\def\vwr{\mathbf{w}^{\rm{r}}}
\def\vfm{\mathbf{\overline{f}}}
\def\vfr{\mathbf{f}^{\rm{r}}}

\def\tr{\mathbf{t}^{\rm{r}}}
\def\tc{\mathbf{t}^{\rm{c}}}
\def\CR{\mathrm{CR}}
\def\QS{\mathrm{QS}}
\def\QSA{\mathrm{QS_a}}
\def\QSB{\mathrm{QS_b}}

\def\PRD{\mathrm{PRD}}
\def\PRDO{\mathrm{PRD_0}}
\def\PRDU{\mathrm{PRD_B}}
\def\PRDN{\mathrm{PRDN}}
\def\tol{\mathrm{tol}}
\def\std{\mathrm{std}}
\def\prd{{\mathrm{prd}}}
\def\mprd{\overline{\mathrm{prd}}}

\def\cumsum{\tt{cumsum}}
\title{Effective high compression 
 of ECG signals at low level distortion} 

\author[1,*]{Laura Rebollo-Neira}
\affil[1]{Mathematics Department,
Aston University, B4 7ET Birmingham, UK}

\affil[*]{l.rebollo-neira@aston.ac.uk}

\begin{abstract}
An effective method for compression of ECG signals, 
which falls within the transform lossy 
compression category,  
is proposed. The transformation is realized by 
 a fast wavelet transform. 
The effectiveness of the approach, in relation to 
the simplicity and speed of its implementation, 
is a consequence of the efficient storage of the 
 outputs of the algorithm which is realized in 
compressed Hierarchical Data Format. 
The compression performance is tested on the 
MIT-BIH Arrhythmia database producing compression 
 results which largely improve upon recently reported 
benchmarks on the same database. For a distortion 
corresponding to a
percentage root-mean-square difference ($\PRD$) of  
 0.53, in mean value, 
 the achieved average compression ratio is 23.17 with quality
score of 43.93. For a mean value of $\PRD$ up to 
1.71 the compression ratio increases up to 62.5.
The compression of a 30 min record is realized 
 in an average time of 0.14 s. The insignificant delay for  
 the compression process, together with the high compression 
ratio achieved at low level distortion and the negligible 
time for the signal recovery, uphold the suitability 
of the technique for supporting distant clinical health care.
\end{abstract}
\begin{document}

\flushbottom
\maketitle
\thispagestyle{empty}

\section*{Introduction}
The electrocardiogram, frequently called ECG, is a 
  routine diagnostic test to assess the electrical and 
muscular functions of the heart. A trained person looking  
at an ECG record can for instance interpret the rate and 
rhythm of heartbeats; estimate the size of the heart, 
 the health of its muscles and its electrical systems; check for effects or side effects of medications on the heart, or check heart abnormalities caused by other health conditions. 
At the present time, ambulatory ECG monitoring serves to detect and characterize abnormal cardiac functions during long hours of ordinary daily activities. Thereby the validated diagnostic role of ECG recording has been extended 
beyond the bedside \cite{GCS07,MMS11,SVC17}. 

The broad use of ECG records, in particular as a 
 mean of supporting clinical health care from a  
distance, enhances the significance of 
dedicated techniques for compressing this type of data. 
Compression of ECG signals may be realized without 
any loss in the signal reconstruction, 
 what is referred to as lossless compression,  
 or allowing some distortion 
 which does not change the clinical information of the 
data. The latter is called lossy compression. This procedure 
can enclose an ECG signal within a file
significantly smaller than that
 containing the uncompressed record. 
  
The literature concerning both 
lossless \cite{JHS90,SE08,SDR11, MMM11,HF17}
and lossy compression \cite{MYL02,KHW10,LKL11,MZD15,FF16,TZW18,EMW17} of ECG records is 
vast. It includes emerging methodologies based on 
compressed sensing \cite{MKA11,ZJM13,PCBV15,PP18}. 
 This work focusses on lossy compression 
 with good performance at low distortion recovery. 
 Even if the approach falls within the standard transform 
 compression category, it achieves stunning results. 
Fresh  benchmarks on the MIT-BIH Arrhythmia database  
 are produced for values of $\PRD$ as 
 in  recent publications \cite{LKL11,MZD15,TZW18,EMW17}.  
 
The transformation step applies 
a Discrete Wavelet Transform (DWT). It is 
recommended to use the fast Cohen-Daubechies-Feauveau 9/7 (CDF 9/7) 
DWT \cite{CWF92}, but other possibilities could also 
 be applied. 
Techniques for ECG signal compression 
 using a wavelet transform
have been reported in numerous publications. For a 
review paper with extensive references see 
 \cite{MD14}. 
 The main difference introduced by our proposal
lies in the compression method.
In particular in what we refer to as the 
Organization and Storage stage. One of the findings of 
this work 
is the appreciation that remarkable compression results
are achievable even prescinding from the typical entropy 
coding step for saving the outputs of the algorithm. 
 High compression is attained in straightforward manner 
 by saving in the Hierarchical Data Format (HDF) 
 \cite{HDF}. More precisely, in the compressed HDF5 version 
which is 
supported by a number of commercial and non-commercial
software platforms including MATLAB, Octave,
Mathematica, and Python. HDF5 also implements a
high-level Application Programming Interface (API) with C,
C++, Fortran 90, and Java interfaces.
As will be illustrated here, if implemented in software, 
adding to the algorithm an entropy coding  process 
may improve compression further, but at expense of 
  processing time. 
 Either way, the compression 
results for distortion corresponding to 
mean $\PRD$ in the range $[0.48, 1.71]$ are shown 
to significantly 
improve recently reported benchmarks \cite{LKL11,MZD15,TZW18,EMW17} on the MIT-BIH Arrhythmia database. 
For $\PRD <0.4$ the technique becomes less effective.

\section*{Method}
\label{Mehtod}
Before describing the method let's introduce the
notational convention. $\R$ is the set of real numbers.
Bold face lower cases are used to represent
one dimension arrays and standard mathematical fonts
to indicate their components, e.g. $\vc \in \R^N$
is an array of $N$ real components $c(i),\, i=1,\ldots,N$,
or equivalently $\vc=(c(1),\ldots,c(N)).$
Within the algorithms, operations on components will be
indicated with a dot, e.g.
 $\vc.^2=(c(1)^2,\ldots, c(N)^2)$ and
$|\vc.| =(|c(1)|, \ldots, |c(N)|)$.
Moreover
$\vt=${\cumsum}$(|\vc.|^2)$ is a vector of
components $t(n)=\sum_{i=1}^n |c(i)|^2, \, n=1,\ldots,N$.

The proposed compression algorithm consists
of three distinctive steps.
\begin{itemize}
\item [1)]
{\em{Approximation Step.}} Applies a  DWT
 to the signal keeping the largest coefficients
 to produce an approximation of the signal up to
 the target quality.
\item [2)]{\em{Quantization Step}}. Uses a scalar quantizer
to convert the  wavelet coefficients
in multiples of integer numbers.
\item [3)]{\em{Organization and Storage Step}}.
 Organizes the outputs of steps 1) and 2)
 for economizing storage space.
\end{itemize}
At the Approximation
Step a DWT is applied to convert the
signal $\vf \in \R^N$ into the vector
$\vw \in \R^N$ whose components are
the wavelet coefficients
$(w(1),\ldots,w(N))$.
For deciding on the number of nonzero coefficients
 to be involved in the approximation we consider
 two possibilities:
\begin{itemize}
\item [a)]The wavelet coefficients
$(w(1),\ldots,w(N))$
 are sorted in
ascending order of their absolute value
$(w(\ga_1),\ldots,w(\ga_N))$,  with
 $|w(\ga_1)|\le \cdots \le |w(\ga_N)|$.
The cumulative sums
$t(n) = \sum_{i=1}^k |w(\ga_i)|^2,\,n=1,\ldots,N$
  are calculated to find all the  values $n$ 
such that $t(n) \ge \tol^2$. Let $k+1$ be the smallest 
of these values. Then the 
indices $\ga_i,\, i=k+1,\ldots N$ give
the coefficients $w(\ga_i),\,i=k+1,\ldots, N$ of  
largest absolute value.
 Algorithm~ \ref{NLW}
summarizes the procedure.
\item [b)]
After the quantization step the
nonzero coefficients and their corresponding indices are
gathered together.
\end{itemize}
\newcounter{myalg}
\begin{algorithm}[!htp]
\refstepcounter{myalg}
\begin{algorithmic}
\caption{Selection of the largest
wavelet coefficients
\newline
Procedure $[\vc, \vll]=\text{SLWC}(\vw,\tol)$}
\label{NLW}
\STATE{{\bf{Input:}}\, Array $\vw$ of wavelets coefficients.
 Parameter $\tol$ for the approximation error.
}
\STATE{{\bf{Output:}}\, Indices $\ell_i,\, i=1,\ldots,K$
of the  selected wavelet coefficients.
Array $\vc$ with the selected wavelet coefficients.}
\STATE{\COMMENT{Sort in ascending order the absolute value
of the components of $\vw$}}
\STATE{$[\vw^{\uparrow},\vga]$={\tt{sort}}($|(\vw.|$)};
\STATE{$\vt=$\cumsum$(\vw^{\uparrow}.^2)$}
\STATE{$\Gamma=$ Set with the values of $n$ such that
 $t(n) \ge \tol^2,\, n=1,\ldots,N$}
\STATE{$\vll=\vga(\Gamma)$}
\STATE{$\vc=\vw(\vll)$}
\end{algorithmic}
\end{algorithm}

At the Quantization Step the  selected
wavelet coefficients
$\vc=(c(1),\ldots,c(K))$, with $K=N-k$ and 
$c(i-k)=w(\ga_i),\, i=k+1,\ldots,N$, are transformed
 into integers by a mid-tread uniform quantizer as follows:
\be
\label{uniq}
c^\Delta(i)= \lfloor \frac{c(i)}{\Delta} +\frac{1}{2} \rfloor,\quad  i=1,\ldots,K.
\ee
where $\lfloor x \rfloor$ indicates the largest
integer number
smaller or equal to $x$ and  $\Delta$ is the quantization
parameter.
After quantization, the  coefficients and indices
are further reduced by the elimination of those coefficients
which are mapped to zero by the quantizer.
The above mentioned option b) follows from
this process. It comes into effect by
skipping Algorithm~\ref{NLW}.
The signs of the coefficients are encoded separately using a binary alphabet (1 for + and 0 for -) in an array
$(s(1),\ldots,s(K))$.

Since the indices ${\ell_i},\,\,i=1,\ldots,K$ are large
numbers, in order to store them in an effective manner
at the Organization and Storage Step we
proceed as follows. These indices
 are re-ordered in ascending
 order $\ell_{i} \rightarrow \til{\ell}_i,\,i=1,\ldots,K$,
which guarantees that
$\til{\ell}_i < \til{\ell}_{i+1},\,i=1,\ldots,K$.
This induces
a re-order in the coefficients,
$\vc^\Delta \rightarrow \vct^\Delta$ and
in the corresponding signs $\vs \rightarrow
\vst$.
The  re-ordered indices are
stored as smaller positive numbers by taking differences
between two consecutive values.
Defining $\delta(i)=\ellt_i-\ellt_{i-1},\,i=2,\ldots,K$
the array
$\vdelt=(\ellt_1, \delta(2), \ldots, \delta(K))$ stores the indices
$\ellt_1, \ldots, \ellt_K$ with unique recovery.
The size of the signal, $N$, the quantization parameter
 $\Delta$, and the
arrays $\vct^\Delta$, $\vst$, and $\vdelt$
are saved in HDF5 format. 
 The HDF5 library 
operates using a chunked storage mechanism. 
The data array is split into equally sized chunks
 each of which is stored separately in the file.
Compression is applied to each individual chunk using
{\tt{gzip}}. The gzip method is based of on the
DEFLATE algorithm, which is a combination of LZ77
\cite{ZL77} and Huffman coding \cite{Huf52}. 
Within MATLAB all this is implemented simply by 
using the function {\tt{save}} to store the data.

Algorithm~\ref{Enc} outlines a pseudo code of
 the above described compression procedure.

\begin{algorithm}[!htp]
\refstepcounter{myalg}
\begin{algorithmic}
\caption{Compression Procedure}
\label{Enc}
\STATE{{\bf{Input:}}\, Array $\vf$ with
the $N$-dimensional signal. Quantization parameter $\Delta$.
Variable case (`a' for
implementation of approach a) and `b' for approach b)).
Parameter $\tol$ for the approximation error.}
\STATE{{\bf{Output:}}\, Array $\vcqt$ with the
quantized re-ordered unsigned wavelet coefficients.
Array $\vdelt$ with the difference of the
 re-ordered indices. Array $\vst$ with the
signs of the re-ordered wavelet coefficients. All these
arrays saved in HDF5 format.}
\STATE{\COMMENT{{\em{Approximation Stage}}}}
\STATE{\COMMENT{Perform a DWT transform of $\vf$ with decomposition level lv}}
\STATE{$N=$numeber of components of $\vf$}
\STATE{$\vw=${{DWT}}($\vf,$lv)}
\IF{case=`a'}
\STATE{\COMMENT{Apply Algorithm~\ref{NLW}}}
\STATE{$[\vc, \vll]=\text{SLWC}(\vw,\tol)$}
\ENDIF
\IF{case=`b'}
\STATE{$\vc=\vw$}
\STATE{$\vll=1: N$}
\ENDIF
\STATE{\COMMENT{{\em{Quantization}}}}
\STATE{Quantize the coefficients $\vc$ to obtain
$\vcq$ from  \eqref{uniq}}
\STATE{Eliminate the zero coefficients and the
corresponding indices}
\STATE{\COMMENT{\em{Organization and Storage}}}
\STATE{\COMMENT{Sort the elements of $\vll$ in ascending order}}
\STATE{$[\vllt,\vgat]$={\tt{sort}}($\vll$)}
\STATE{\COMMENT{Set $\til{\delta}(1)=\ellt_1$ and take the
difference of the re-ordered indices}}
\STATE{$\til{\delta}(i)=\ellt_i-\ellt_{i-1},\,i=2,\ldots,K$}
\STATE{$\vcqt=|\vcq(\vgat).|$}
\STATE{$\vst=${(\tt{sign}}$(\vcq(\vgat))+ 1)./2$}
\STATE{Save the number $N$
and the arrays $\vcqt$,
$\vdelt$ and $\vst$ in HDF5}
\end{algorithmic}
\end{algorithm}

The fast wavelet transform
has computational complexity O($N$). Thus,
if the approach a) is applied,
the computational complexity of Algorithm \ref{Enc}
 is dominated by the
 sort operation in Algorithm \ref{NLW}
with average computational complexity
$\text{O}(N\log N)$.
 Otherwise the complexity is just $\text{O}(N)$, because
the number $K$ of indices of nonzero coefficients  to
be sorted is in general much less than $N$. 
Nevertheless, as seen in 
Tables~\ref{TABLE3}-\ref{TABLE6}, 
in either case the compression of
 a 30 min record is achieved on a MATLAB platform in 
 an average time less then 0.2 s. 
While compression performance can be 
improved further by adding an entropy coding 
step before saving the arrays, if implemented in 
software such a step slows the process. 

When selecting the number of wavelet coefficients for the
approximation by method
a) the parameter $\tol$ is fixed as follows:
Assuming that the target $\PRD$ before quantization is
$\PRDO$ we set $\tol = \PRDO \|f\|/100$.  The 
value of $\PRDO$ is fixed as 70\%- 80\%  of the 
required $\PRD$. The quantization parameter is 
tuned to achieve the required $\PRD$.

\subsection*{Signal Recovery}

At the Decoding Stage the signal is recovered by the
following steps.
\begin{itemize}
\item
Read the number $N$,
the quantization parameter $\Delta$, and
the arrays $\vcqt$, $\vdelt$, and $\vst$ from the compressed file.
\item
Recover the magnitude of the coefficients
from their quantized version as
\be
\til{\vc}^{\mathrm{r}}=
\Delta \,{\til{\vc}}^\Delta.
\ee
\item Recover the indices $\vllt$
 from the array $\vdelt$ as: $\ellt_1= \til{\delta}(1)$
and $\ellt_i=\til{\delta}(i)+\til{\delta}(i-1),\,i=2,\ldots,K.$
\item
Recover the signs of the the wavelet coefficients as
$\vstr=2\,\vst-1$
\item
Complete the full array of wavelet coefficients as
$\vwr(i)=0,\,i=1,\ldots,N$ and
$\vwr(\vllt)= \vstr. \til{\vc}^{\mathrm{r}}$
\item
Invert the wavelet transform to
recover the approximated signal $\vfr$.
\end{itemize}
As shown in Tables~\ref{TABLE3}-\ref{TABLE5}, and 
 the recovery process runs
 about 3 times faster than the compression procedure,
which is already very fast.


\section*{Results}
\label{Results}
We present here four numerical tests with
 different purposes.  Except for the comparison in 
 Test II, all the other tests use the
 full MIT-BIH Arrhythmia database \cite{MITDB}
which contains 48 ECG records. Each of these records 
consists of $N=650000$ 11-bit samples at a frequency 
of 360 Hz. The algorithms are implemented using MATLAB in a
notebook Core i7 3520M, 4GB RAM.

Since the compression performance of lossy compression 
has to be considered in relation to the quality of 
the recovered signals, we introduce at this 
point the measures to evaluate the results of 
the proposed procedure.

The quality of a recovered signal is assessed
 with respect to the $\PRD$
calculated as follows,
\be
\PRD=\frac{\|\vf - \vfr\|}{\|\vf\|} \times 100 \%,\quad
\ee
where, $\vf$ is the original signal, $\vfr$ is
the signal reconstructed from the compressed file and
$\| \cdot\|$ indicates the 2-norm. Since the
$\PRD$  strongly depends on the baseline of the signal,
the $\PRDN$, as defined below,
 is also reported.
\be
\PRDN=\frac{\|\vf - \vfr\|}{\|\vf - \vfm\|} \times 100 \%,\quad
\ee
where, $\vfm$ indicates the mean value of $\vf$.

When fixing a value of $\PRD$, the compression performance
is assessed by the Compression Ratio (CR) as given by
\be
\text{CR}=\frac{{\text{Size of the uncompressed file}}.}{{
\text{Size of the compressed file}}}
\ee
The quality score ($\QS$),
reflecting the tradeoff between
compression performance and reconstruction quality,
is the ratio:
\be
\QS=\frac{\CR}{\PRD}.
\ee
Since the $\PRD$ is a global quantity, in order to
detect possible local changes in the visual quality
of the recovered signal,
 we define the local $\PRD$ as follows.
Each signal is partitioned in $Q$ segments
$\vf_q,\, q=1\ldots,Q$ of $L$ samples.
 The local $\PRD$ with respect to every segment
in the partition, which we indicate 
as $\prd(q),\,q=1,\ldots Q$, is calculated as
\be
\prd(q)=\frac{\|\vf_q - \vfr_q\|}{\|\vf_q\|} \times 100 \%,\quad
\ee
where $\vfr_q$ is the recovered portion
of the signal corresponding to the segment $q$.
For each record the mean value $\prd$ ($\mprd$)
 and corresponding standard deviation ($\std$)
   are calculated as
\be
\mprd=\frac{1}{Q}\sum_{q=1}^Q \prd(q)
\ee
and
\be
\std= \sqrt{\frac{1}{Q-1}\sum_{q=1}^Q (\prd(q)
- \mprd)^2}.
\ee
The mean value $\prd$ with
 respect to all the records in the
 database is a double average $\overline{\mprd}$.

When comparing two approaches on a database we reproduce the 
same mean value PRD. The quantification of the 
relative gain in CR of one particular approach, say  approach 1, in relation to another, say  approach 2, is  given by the quantity:
$$\text{Gain}=\frac{\CR_1 - \CR_2}{\CR_2} \times 100 \%.$$
The gain in QS has the equivalent definition.

\subsection*{Numerical Test I}
We start the tests by implementing the
proposed approach using wavelet transforms
corresponding to different wavelet families  at
  different levels of decomposition.
The  comparison between different wavelet 
 transforms is realized using approach b),
because within this option each value of
 $\PRD$ is uniquely determined by the quantization
parameter $\Delta$. Thus, the difference in
$\CR$ is only due to the particular wavelet basis
and the concomitant decomposition level.
Table~\ref{family} shows the average $\CR$ (indicated as 
$\CRB$) and
corresponding standard deviation (std) with
respect to the whole data set and for
three different values of
 $\PRD$.  For each  $\PRD$-value
 $\CRB$  is obtained by means of
the following wavelet basis: db5 (Daubechies) 
coif4 (Coiflets)
sym4 (Symlets) and cdf97 (Cohen-Daubechies-Feauveau).
Each basis is decomposed in three
different levels (lv).

As observed in Table~\ref{family}, 
on the whole the best $\CR$ is
achieved with the biorthogonal basis cdb97
for lv=4. In what follows 
all the results are given using this basis 
for decomposition level lv=4.

\begin{table}[h]
\caption{Comparison of CRs for 
three values of $\PRD$  when the proposed 
approach is implemented using different 
wavelets at decomposition levels 3, 4, and 5.}
\label{family}
\begin{center}
\begin{tabular}{||c|c|c|c|c|c|c|c|c|c|c|c|c|c||}\hline \hline
Family&lv& $\Delta$ &PRD&CR&std&$\Delta$&PRD&CR&std&$\Delta$&PRD&CR&std\\\hline
\multirow{3}{*}   
 &3&47&0.65&24.03&5.44&36&0.53&20.25&4.54&29&0.45&{\bf{17.53}}&3.95\\ 
\cline{2-14}
db5&4&52&0.65&{\bf{25.12}}&6.07&39&0.53&{\bf{20.75}}&4.97&
30 &0.45&{{17.37}}&4.08\\ \cline{2-14}
&5&53&0.65&23.31&5.51&39&0.53&19.22&4.48&30&0.45&15.49&3.84\\
\hline \hline
\multirow{3}{*}  
 &3&47&0.65&24.28&5.59&36&0.53&20.47&4.56&29&0.45& 17.76&3.98\\
 \cline{2-14}
coif4&4&52&0.65&{\bf{25.56}}&6.35& 39 &0.53&{\bf{21.15}}&5.09&31 &0.45& {\bf{18.10}}&4.28\\  \cline{2-14}
&5&53&0.65&23.73&5.75&39&0.53&19.61&4.59&31&0.45& 16.39&4.17\\
\hline \hline
\multirow{3}{*} 
 &3&47&0.65&23.65&5.30&36&0.53&19.95&4.43&28 &0.45& 17.01&3.78\\
\cline{2-14}
sym4&4&52&0.65&{\bf{25.13}}&6.16& 39 &0.53&{\bf{20.76}}&4.91&30 &0.45& {\bf{17.41}}.&4.10\\ \cline{2-14}
&5&53&0.65&23.64&5.75& 39&0.53&19.42&4.50&30&0.45& 15.84&3.98\\
\hline \hline
\multirow{3}{*} 
 &3&47&0.65&24.66&5.39&36&0.53&20.94&4.66&28&0.45&17.80&3.93\\
\cline{2-14}
cdf97&4&52 &0.65&{\bf{26.59}}&6.42&39&0.53&{\bf{22.16}}&5.27&30&0.45&{\bf{18.57}}&4.39\\ \cline{2-14}
&5&53&0.65&24.98&6.06&39&0.53&20.69&4.83&30&0.45&16.75&4.20\\\hline \hline
\end{tabular}
\end{center}
\end{table}

Next
we produce the $\CR$ for every record in the database
 for a mean value $\PRD$ of 0.53.

\begin{table}[H]
\caption{Compression results with approach a), 
 cdf97 DWT,  lv=4,  
 $\Delta=35$, and $\PRDO=0.4217$, 
for the 48 records in the MIT-BIH Arrhythmia Database
listed in the first column of the left and right parts of the table.}
\label{TABLE1}
\begin{center}
\begin{tabular}{||r|r|r|r|r|r|r||}
\hline \hline
Rec &$\mprd$&$\std$&$\PRD$ &$\CRA$&$\QSA$&$\PRDN$
\\ \hline \hline
100&  0.52&  0.02& 0.52& 28.65&  55.01&   12.99\\ 
101&  0.51&  0.08& 0.52& 28.32&  54.92&   9.56 \\
102&  0.52&  0.03& 0.52& 29.15&  55.89&   13.36 \\
103&  0.52&  0.04& 0.52& 26.32&  50.86&   7.88 \\
104&  0.52&  0.12& 0.53& 21.23&  40.08&   10.37 \\
105&  0.52&  0.06& 0.53& 20.07&  38.08&   6.39 \\
106&  0.51&  0.07& 0.52& 20.46&  39.49&   6.97 \\
107&  0.54&  0.05& 0.54& 14.30&  26.66&   3.10 \\
108&  0.52&  0.09& 0.52& 22.52&  43.11&   8.51 \\
109&  0.52&  0.05& 0.52& 23.80&  45.42&   5.16 \\
111&  0.52&  0.04& 0.52& 26.71&  51.55&   10.01 \\
112&  0.54&  0.06& 0.55& 28.11&  51.45&   10.52 \\
113&  0.52&  0.02& 0.52& 22.89&  43.93&   6.29 \\
114&  0.51&  0.04& 0.52& 31.85&  61.80&   14.94 \\
115&  0.53&  0.03& 0.53& 22.02&  41.68&   6.75 \\
116&  0.58&  0.04& 0.58& 12.84&  22.05&   3.71 \\
117&  0.54&  0.03& 0.54& 33.70&  62.86&   9.59 \\
118&  0.61&  0.07& 0.62& 12.11&  19.69&   6.16 \\
119&  0.55&  0.02& 0.55& 18.01&  32.67&   4.42 \\
121&  0.53&  0.06& 0.53& 38.74&  73.18&   7.59 \\
122&  0.55&  0.02& 0.55& 21.36&  38.58&   6.49 \\
123&  0.54&  0.03& 0.54& 28.08&  52.05&   8.05 \\
124&  0.54&  0.05& 0.54& 26.03&  48.21&   5.07 \\
200&  0.52&  0.07& 0.53& 16.51&  31.19&   7.00 \\
201&  0.51&  0.05& 0.52& 37.62&  72.79&   13.23\\
\bottomrule
\end{tabular}
\begin{tabular}{||r|r|r|r|r|r|r||}
\hline \hline
Rec &$\mprd$&$\std$&$\PRD$&$\CRA$&$\QSA$&$\PRDN$
\\ \hline \hline
202&  0.51&  0.05& 0.51& 30.41&  59.57&   8.51\\
203&  0.54&  0.08& 0.54& 13.64&  25.11&   5.46\\
205&  0.52&  0.03& 0.52& 30.27&  57.77&   12.83\\
207&  0.50&  0.11& 0.52& 30.31&  58.72&   7.23\\
208&  0.52&  0.07& 0.53& 15.98&  30.38&   5.43\\
209&  0.52&  0.07& 0.53& 16.43&  31.08&   9.79\\
210&  0.51&  0.09& 0.51& 26.30&  51.08&   9.80\\
212&  0.54&  0.08& 0.54& 13.28&  24.37&   8.18\\
213&  0.54&  0.03& 0.54& 13.60&  25.09&   3.99\\
214&  0.52&  0.05& 0.52& 21.45&  41.45&   5.48\\
215&  0.54&  0.06& 0.54& 15.10&  27.95&   9.53\\
217&  0.52&  0.03& 0.52& 18.13&  34.83&   4.22\\
219&  0.54&  0.03& 0.54& 18.69&  34.44&   4.50\\
220&  0.54&  0.03& 0.54& 24.21&  44.77&   7.79\\
221&  0.51&  0.04& 0.51& 24.05&  46.93&   8.46\\
222&  0.51&  0.07& 0.52& 24.48&  47.17&   13.88\\
223&  0.54&  0.03& 0.54& 22.24&  41.46&   6.10\\
228&  0.52&  0.08& 0.52& 19.23&  36.91&   7.51\\
230&  0.52&  0.06& 0.52& 21.36&  41.04&   7.28\\
231&  0.52&  0.04& 0.52& 27.10&  51.81&   9.56\\
232&  0.51&  0.07& 0.51& 34.34&  66.73&   15.50\\
233&  0.53&  0.05& 0.53& 15.74&  29.59&   4.89\\
234&  0.52&  0.03& 0.52& 24.47&  47.10&   7.65\\ \hline
{\bf{mean}}& {\bf{0.53}} & {\bf{0.05}} & {\bf{0.53}} & 
{\bf{23.17}} &{\bf{43.93}}& {\bf{8.08}} \\ \hline
{\bf{std}}& {\bf{0.02}}  &{\bf{0.02}} & {\bf{0.02}}  & {\bf{6.67}}& {\bf{13.23}}& {\bf{3.06}} \\ \hline
\bottomrule
\end{tabular}
\end{center}
\end{table}
Table~\ref{TABLE1} shows the results obtained 
by approach a) where the  
  $\CR$ and $\QS$ produced
 by this method are indicated as $\CRA$ and $\QSA$, 
respectively. The $\PRD$ values
 for each of the records listed in the first
column of Table~\ref{TABLE1} 
are given in the forth columns of those tables.
Notice that the order refers to both, the
 left and right parts of the table.
The second and third columns show the values of
$\mprd$ and the corresponding $\std$ for
each record. The $\CR$ is given in the
fifth column and the corresponding $\QS$ in
sixth column of the table.
The mean value $\CR$ obtained
by method b) for the same mean value $\PRD= 0.53$
is $\CRB=22.16$.

Table~\ref{PRD0} shows the variations of the 
$\CRA$ with different values of the parameter $\PRDO$ in 
method a).

\begin{table}[h]
\caption{Comparison of the CR achieving $\PRD=0.53$ with 
 method a) of the proposed approach  for 
different values of the parameter $\PRDO$.}
\label{PRD0}
\begin{center}
\begin{tabular}{||l||r|r|r|r|r|r||}\hline\hline
$\PRDO\,$ & 0.212& 0.265& 0.318& 0.371&0.424& 0.477\\ \hline
$\CRA$   &22.16 & 22.19&22.42&22.88 &23.24& 19.50 \\ \hline
$\Delta$& 39 & 39& 39 & 37&  32& 21\\ \hline  \hline
\end{tabular}
\end{center}
\end{table}

\subsection*{Numerical Test II}
Here comparisons are carried out with respect to
 results produced by the
set partitioning in hierarchical threes algorithm (SPHIT)
 approach  proposed in \cite{LKP00}.
Thus for this  test we use
 the data set described
in that publication. It consists of 10-min long
  segments from records
 100, 101, 102, 103, 107, 108, 109, 111, 115, 117, 118, and
111. 
As indicated in the footnote of \cite{LKP00}
at pp 853, the given values of PRD
 correspond to the subtraction of a
 baseline equal to 1024.  This has generated confusion in the
literature, as often the values of PRD in Tables III of
\cite{LKP00} are unfairly reproduced for comparison with values of PRD obtained without
subtraction of the 1024 base line.
The values of PRD with and without
subtraction of that baseline, which are  indicated
as $\PRDU$ and $\PRD$ respectively, are given in
Table~\ref{sphit1}.
 As seen in this 
table, for the same approximation there is an
enormous difference between the two metrics. A fair comparison
with respect to the results in \cite{LKP00} should either involve 
the figures
in the second row of Tables~\ref{sphit1} or,
 as done in \cite{LKP00}, the fact that 
 a 1024 base line has been subtracted should be specified.
\begin{table}[h]
\caption{Comparison with the results of Table III in \cite{LKP00}}
\label{sphit1}
\begin{center}
\begin{tabular}{||l||r|r|r|r|r|r|r||}\hline\hline
$\PRDU $ &1.19& 1.56& 2.46& 2.96& 3.57& 4.85& 6.49\\\hline 
$\PRD$  &0.11& 0.15& 0.23& 0.28& 0.35& 0.47& 0.63\\\hline
$\CR$ \cite{LKP00} &  4 &   5  &  8  & 10  &  12 &  16 &  20 \\\hline
$\CRB$&4.16&5.29&8.79&11.24&14.11&19.13&24.64 \\ \hline
Gain \%&  4& 6& 10& 12& 18& 20& 23 \\\hline 
$\CRBH$&5.20&6.57&10.53&13.19&16.04&21.31&26.92\\\hline
Gain \%& 30 & 31 & 32 & 32& 34& 33 & 35 \\\hline
$\Delta$& 3.87&5.45&8.79&14.53&20.10&33.10&51.50\\
\hline \hline
\end{tabular}
\end{center}
\end{table}

The figures in the 3rd row of Tables~\ref{sphit1} 
  correspond to the
 CRs in \cite{LKP00}.
 The 4th 
row  shows the  CRs resulting from method b) of
 the proposed approach without entropy coding and 
the 5th row the results of adding a Huffman coding
step  before saving the compressed data in HDF5 format.
 The last two rows show  the
quantization parameters $\Delta$ which
produce the required values of $\PRDU$ and $\PRD$.
\subsection*{Numerical Test III}
This numerical test aims at comparing our results with
recently reported benchmarks on the full MIT-BIH Arrhythmia
database for mean value PRD in the rage $[0.23, 1.71]$.
 To the best of our knowledge the  highest  CRs
reported so far for mean value PRD
in the range $[0.8, 1.30)$ are those
 in \cite{MZD15}, and
in the range $(1.30, 1.71]$  those in
 \cite{TZW18}. For  $\PRD< 0.8$  the
comparison is realized with the results in
\cite{LKL11}, as shown in Table~\ref{TABLE5}.
Table~\ref{TABLE3}
compares our results 
 against the results in
Table III of \cite{MZD15} and Table~\ref{TABLE4}
 against
Table 1 of \cite{TZW18}. In both cases we reproduce the
identical mean value of PRD. The differences are in the
values of CR and QS.
All the Gains given in Table~\ref{TABLE3} are relative to
the results
in \cite{MZD15} while those given in Table~\ref{TABLE4}
and Table~\ref{TABLE5} are relative to the results in
\cite{TZW18} and \cite{LKL11}, respectively.

\begin{table}[h]
\caption{Comparison between the average
performance of the proposed method and the method
in \cite{MZD15} for the same mean value of $\PRD$.}\label{TABLE3}
\begin{center}
\begin{tabular}{||l||r|r|r|r|r|r||}
\hline \hline
$\PRD$ &1.71&1.47&1.18&1.05&0.91&0.80\\
\hline \hline
$\CR$\cite{MZD15}&38.46&33.85&28.21&25.64&22.27&18.00\\  \hline
$\CRA$&{\bf{62.48}}&{\bf{56.78}}&{\bf{47.04}}&{\bf{41.24}}& {\bf{37.68}}&{\bf{33.05}}\\
Gain \% &62&68&67&61&69&84\\ \hline
$\CRB$&60.33&53.07&44.37&40.29&35.69&31.86\\
Gain \%&57&57&57&57&60&77\\ \hline
$\QS$\cite{MZD15}&29.08& 29.38&30.01&30.51&30.36&29.46\\ \hline
$\QSA$&{\bf{36.55}}&{\bf{38.63}}&{\bf{39.83}}&{\bf{39.20}}&{\bf{41.61}}&{\bf{41.55}}\\
Gain \% &26&31&33&28&37&41\\ \hline
$\QSB$&35.86&36.81 &38.33&39.16&39.96& 40.80\\
Gain \% &23&25&27&28&31&38\\
\hline
$\tc$ (s) a)&0.13 & 0.14 &0.14 &0.14&0.14&0.14\\
$\tc$ (s) b)&0.10 & 0.11 &0.11 &0.11&0.11&0.11\\
$\tr$ (s)&0.04 & 0.05 &0.05 &0.05 &0.05&0.05\\
\hline \hline 
$\PRDO$  a)&1.303& 0.966 & 0.830 & 0.830 & 0.635 &0.627\\
$\Delta\,$ a)&119& 126 & 93 & 67 & 71 & 51\\
$\Delta\,$ b)& 177 & 147 & 113 & 98 & 82 &69\\
\hline \hline
\end{tabular}
\end{center}
\end{table}

As already remarked, and fully discussed in  \cite{BVCRGL05},
 when comparing results from different publications
 care should be taken to make sure that the comparison is
actually on the identical database,
 without any difference in baseline. From the
information given in the papers producing the
results we are comparing with (the relation between
the values of PRD and PRDN)
 we can be certain that we are  working on
 the same database \cite{MITDB},
which is described in \cite{MM01}.

The parameters for reproducing the
required $\PRD$ with methods a) and b) are
given in the last 3 rows of Tables~\ref{TABLE3}-\ref{TABLE5} The previous 3 rows in
each table give, in seconds, the average time to
compress ($\tc$) and recover ($\tr$) a record.
As can be observed, the compression times of approaches
a) and b) are very similar.
The  given times were obtained as the average of
 10 independent runs.  Notice that the $\CR$ in
these tables do not include the additional
entropy coding step.


\begin{table}[h]
\caption{Comparison between the average
performance of the proposed method and the method
in \cite{TZW18} for the same mean value of $\PRD$.}
\label{TABLE4}
\begin{center}
\begin{tabular}{||l||r|r|r|r|r|r||}
\hline \hline
$\PRD$ &1.71&1.47&1.29&1.14&1.03&0.94\\
\hline \hline
$\CR$ \cite{TZW18}&42.27&35.53&30.21& 25.99&22.80&20.38\\  \hline
$\CRA$& {\bf{62.48}}&{\bf{56.78}}&{\bf{49.60}}&{\bf{45.75}}&{\bf{41.00}}&{\bf{38.52}}\\
Gain \%& 48 & 60  &64 & 76 & 80 & 89 \\  \hline
$\CRB$&60.33&53.07&48.04&42.94&39.76&36.52\\
Gain \%&43& 49& 59& 65& 74& 79\\ \hline
$\QS$ \cite{TZW18}&33.41& 32.58&31.53&30.23&29.19&28.44\\ \hline
$\QSA$&{\bf{36.55}}&{\bf{38.63}}&{\bf{38.43}}&{\bf{40.28}}&{\bf{39.72}}&{\bf{41.03}}\\
Gain\%& 7& 18& 21& 33& 36& 44\\ \hline
$\QSB$&35.86&36.81&38.00&38.50& 39.26 & 39.77\\
Gain \%& 11 & 15 &19 &24 &29 &33\\\hline
$\tc$ a) (s)&0.14 &0.14 &0.14 &0.14 &0.14 &0.14\\
$\tc$ b) (s)&0.10 &0.11 &0.11 &0.11 &0.11 &0.11\\
$\tr$ (s)&0.05 &0.05 &0.05 &0.05 &0.05 &0.05\\
\hline \hline
$\PRDO$ a)&1.303 & 0.966 & 0.971 & 0.750 &0.804&0.690\\
$\Delta$ a)&119&126&91&96 &68 & 69\\
$\Delta$ b)&177& 147 & 126 & 108 &96 & 85\\
\hline \hline
\end{tabular}
\end{center}
\end{table}

\begin{table}[H]
\caption{Comparison between the average compression
performance of the proposed method and the method
in \cite{LKL11} for the same mean value of $\PRD$.}\label{TABLE5}
\begin{center}
\begin{tabular}{||l||r|r|r|r|r|r||}
\hline \hline
$\PRD$ &1.31&1.02& 0.67 & 0.48&  0.31& 0.23\\
\hline \hline
$\CR$\cite{LKL11}&17.34 & 14.68 & 11.30 & 9.28 & 6.22& 5.19\\
\hline
$\CRA$& {\bf{49.99}} & {\bf{40.57}} & {\bf{27.79}} & {\bf{19.84}} & {\bf{10.59}}&{\bf{7.75}}\\
Gain\% & 188 &{{176}} &{{146}}& {{114}}& {{70}} &{{49}}\\ \hline
$\CRB$&48.72&39.47&27.24& 19.84&10.33&7.36\\
Gain\%&181& 169& 141& 114& 66& 42 \\\hline
$\QSA$&38.12&39.81&41.46& 42.71& 34.96 & 34.48\\
$\QSB$&37.69&39.30&41.88&42.71&34.40&33.16\\
\hline 
$\tc$ (s) a)&0.14 &0.14 &0.14 &0.14 &0.15 &0.16\\
$\tc$ (s) b)&0.11 &0.11 &0.11 &0.11 &0.13 &0.14\\
$\tr$ (s)   &0.05 &0.05 &0.05 &0.05 &0.05&0.06\\
\hline \hline
$\PRDO$  a)&1.000 & 0.794 & 0.562 & 0.380 & 0.224 & 0.193\\
$\Delta$ a)&90 &67 &36 & 33 &16&9\\
$\Delta$ b)&129& 95&54 &33 &16 &10\\ \hline \hline
\end{tabular}
\end{center}
\end{table}

\begin{figure}[H]
\begin{center}
\includegraphics[width=13cm]{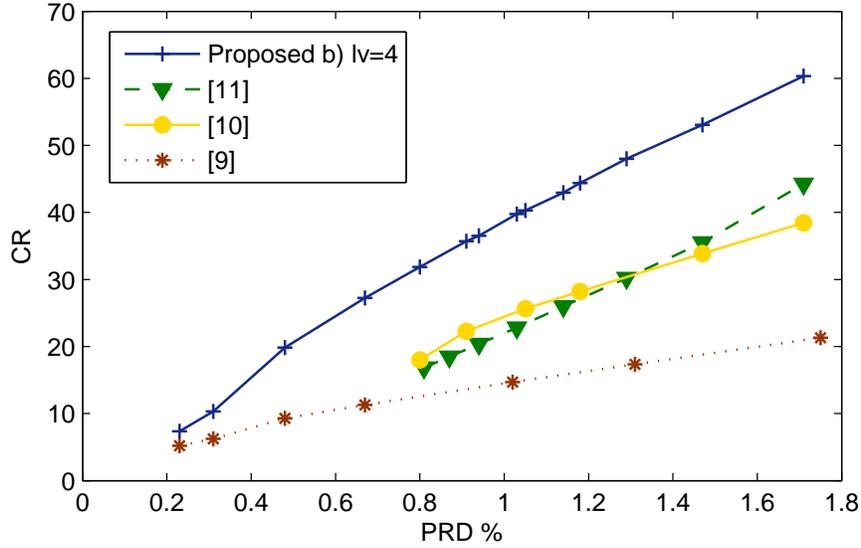}
\caption{$\CR$ vs $\PRD$ corresponding to the
proposed approach method b) (blue line)
and the approaches in \cite{MZD15} (green line),
\cite{TZW18} (yellow line) and \cite{LKL11} red line.}
\label{Comp_Fig}
\end{center}
\end{figure}

Fig.~\ref{Comp_Fig} gives the plot of CR vs PRD for
the approaches being compared in this section.

\subsection*{Numerical Test IV}
Finally we would like to highlight the following two
features of the proposed compression algorithm.
\begin{itemize}
\item [1)] One of the distinctive features stems from
the significance
of saving the outputs of the algorithm directly in 
compressed HDF5 format.
In order to highlight this, we compare the
  size of the file saved in this way 
 against the size of the file obtained by applying 
 a commonly used entropy coding process, Huffman coding, 
 before saving the data in HDF5 format. 
 The implementation of Huffman coding is realized, 
 as in Table~\ref{sphit1}, by the off the shelf 
 MATLAB functions {\tt{Huff06}} available on \cite{Karl}. 
 In  
 Table~\ref{TABLE6} $\CRA$ and $\CRB$ indicate, 
as before, the CR obtained when the outputs of 
methods a) and 
b) are directly saved in HDF5 format. $\CRAH$  and $\CRBH$
 indicate the  CR when Huffman
coding is applied on the outputs a) and b) before
 saving the data in HDF5 format. The rows right 
below the CRs give the corresponding compression times. 
\item [2)] The other distinctive feature of the 
 method is the
 significance of the proposed  Organization
 and Storage step. In order to illustrate this,
 we compare the 
 results obtained by method b) with those 
 obtained using the
 conventional Run-Length (RL) algorithm \cite{Sal07}
 instead of storing the indices of nonzero coefficients as
proposed in this work. The CR corresponding to   
 RL in HDF5 format is indicated in 
Table~\ref{TABLE6} as $\CRRL$. When   
Huffman coding is applied on RL before saving the outputs 
in compressed HDF5 format, the CR is indicated as $\CRRLH$. 
\end{itemize}
\begin{table}[h]
\caption{Comparison of different storage methods. 
$\CRA$  and $\CRB$ are the CRs 
from approaches a) and b) when the outputs are 
saved directly in HFD5 format. $\CRAH$ and 
 $\CRBH$ are the corresponding values when 
the Huffman codding step is applied 
before saving the data in HFD5 format.
$\CRRL$ gives the CR if the outputs of 
method b) are stored using the RL algorithm and 
 the arrays are saved in
HFD5 format. $\CRRLH$ is the corresponding CR if 
Huffman codding is applied before saving 
the arrays in HFD5 format.} 
\label{TABLE6}
\begin{center}
\begin{tabular}{|l||r|r|r|r|r|r|r|r|r||}
\hline \hline
$\PRD$ &1.0&0.9& 0.8 & 0.7& 0.6& 0.5&0.4&0.3&0.2\\
\hline \hline
$\CRA$ &40.51& 37.12 & 33.09 &29.70&25.50& 22.00&16.80&10.33&6.64\\
$\CRAH$&43.57&40.41&36.32& 32.96&28.80&25.13&20.25&14.62&9.53\\
Gain \%& 8 & 9& 10& 11& 13& 14& 20& 42& 43\\ \hline
$\tc$ (s)&0.13 &0.13 &0.13 &0.14 &0.14 &0.14&0.15&1.15&1.15\\
$\tc$ (s)&4.3 &4.5 & 5.0&5.4 &5.5 &6.2& 8.3&10.22&15.4\\
\hline 
$\Delta$ a)&71 &64 &50& 45 &35& 30& 24&15&8.5\\
$\PRDO$ a)&0.750 & 0.675 & 0.640& 0.550 & 0.484 & 0.400 & 0.300& 0.230&0.150\\
\hline \hline
$\CRB$& {{38.64}} & {{35.41}} & {{31.86}} & {{28.53}} & {{24.93}}&{{21.03}}& 16.21& 10.07& 6.56\\
$\CRBH$&42.56 &39.20&  35.65& 32.10& 28.37& 24.32& 19.60& 14.32& 9.40\\
Gain \% & 10 & 11 & 12& 13& 14& 16& 21& 42& 43\\\hline
$\tc$ (s)&0.11 &0.11 &0.11 &0.11 &0.11 &0.11 &0.12 &0.12&0.12\\
$\tc$ (s)&4.1 &4.0& 4.5&5.2& 5.4& 6.7& 8.1&10.7& 15.5\\ \hline
$\Delta$ b)&92&81 &69 & 58 &47&36& 25 & 15.5& 8.5\\
\hline \hline
$\CRRL$&26.63 & 24.41 &  22.42 & 20.34 & 18.14& 15,68 & 12.50& 8.32& 5.61\\
$\CRRLH$& 35.06 & 31.78 & 28.80 & 25.93 & 22.91 & 19.66 
& 15.85 &11.63 & 7.82\\ 
Gain \% &32 & 30& 28& 27 & 26 & 25 & 26& 40 & 40 \\ \hline
$\tc$ (s)&0.12 &0.12 &0.12 &0.13 &0.13&0.13& 0.13& 0.14& 0.16\\
$\tc$ (s)&4.5 & 4.9 &6.2 &6.4& 7.1& 7.4&9.0&12.8&19.5\\
\hline \hline
\end{tabular}
\end{center}
\end{table}
\section*{Discussion}

We notice that, while the results in Table~\ref{family}
 show some differences in CR when different 
 wavelets are used for the DWT, 
it is clear from the table that the selection of the 
wavelet family is not the crucial factor for 
the success of the technique. The same is 
true for the decomposition level. That said, 
since the best results correspond to the 
cdf97 family at decomposition level 4, we have realized
 the other numerical tests with that wavelet basis. 

\begin{figure}[ht]
\begin{center}
\includegraphics[width=11cm]{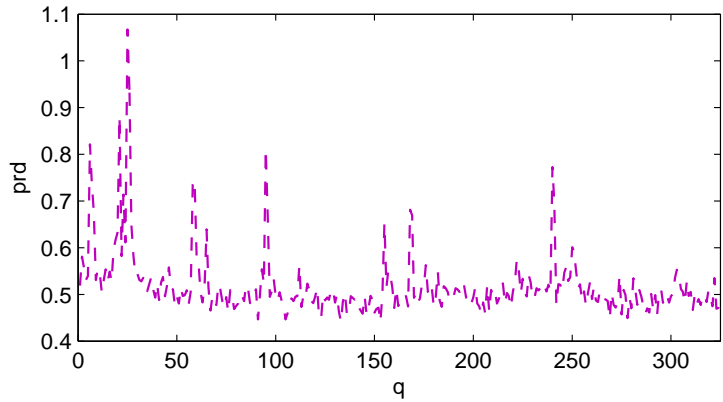}\\
\end{center}
\caption{Values of $\prd$ for the $Q=325$
 segments of length $L=2000$ in record 101.}
\label{prds}
\end{figure}

\begin{figure}[H]
\begin{center}
\includegraphics[width=8.7cm]{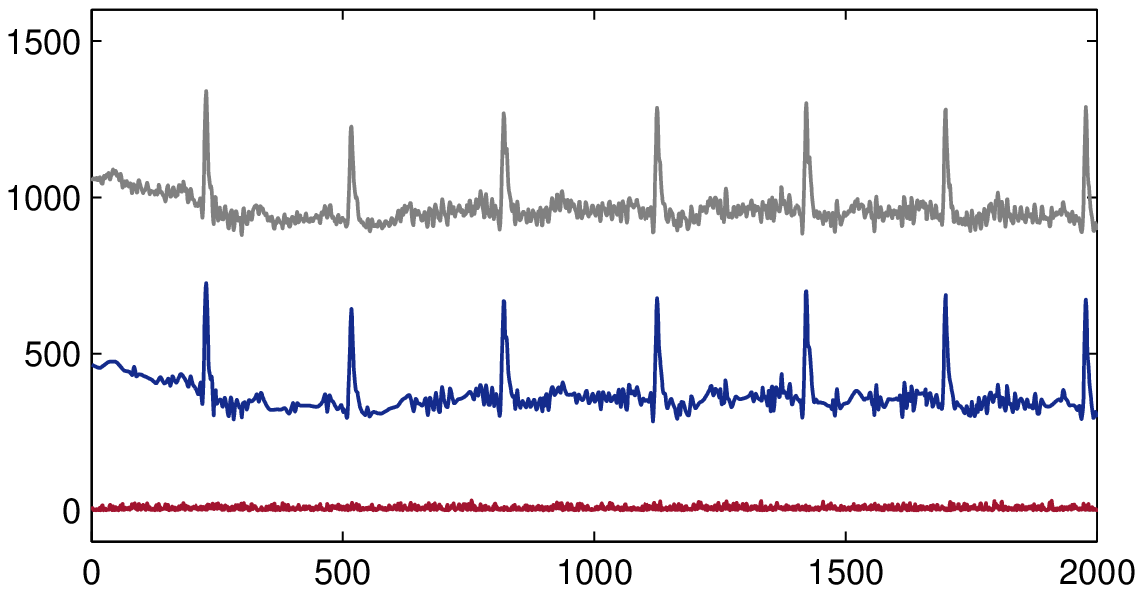}
\includegraphics[width=8.7cm]{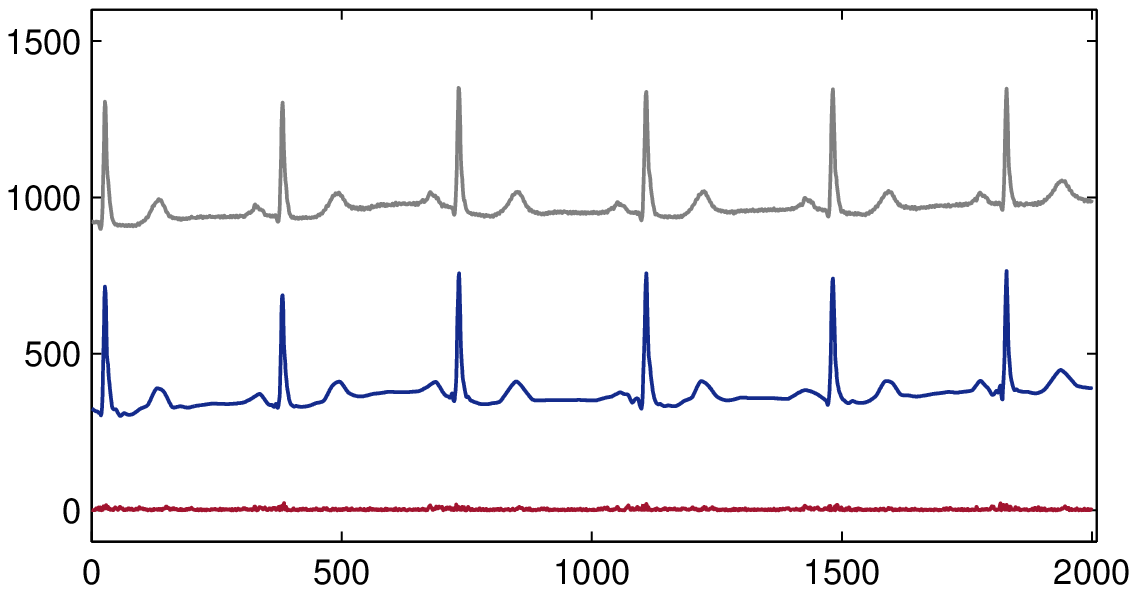}
\end{center}
\caption{The upper waveforms in both graphs
 are the raw data.
 The lower waveforms are the corresponding
approximations which have been shifted down for visual convenience. The bottom lines  represent the absolute value of
the difference between the raw data and the approximation.
The left graph corresponds to  segment 25 in record 101 and
the right graph corresponds to  segment 120 in the same
record.}
\label{Rec1}
\end{figure}

\begin{figure}[H]
\begin{center}
\includegraphics[width=8.7cm]{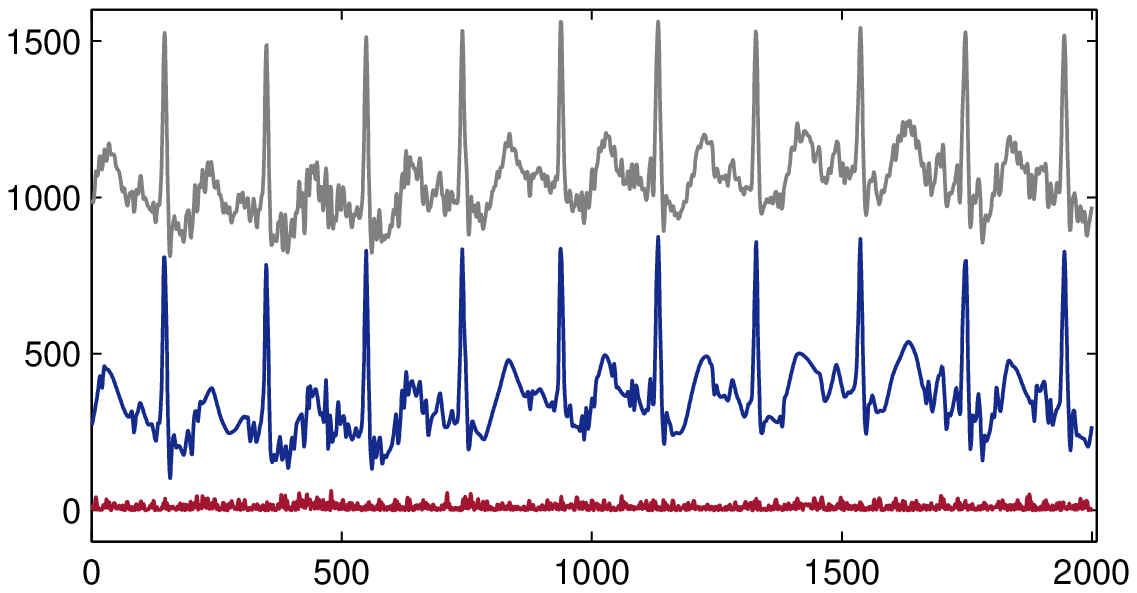}
\includegraphics[width=8.7cm]{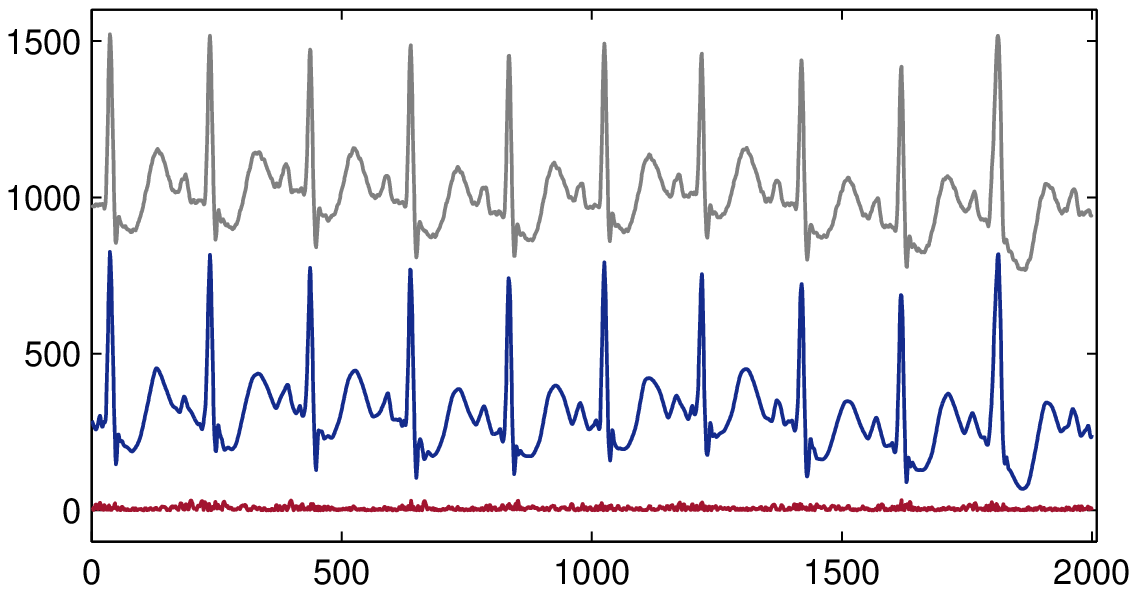}
\end{center}
\caption{Same description as in Fig.~\ref{Rec1} but
for record 213 and segment 175 (left graph) and
51 (right graph).}
\label{Rec2}
\end{figure}

We chose to produce full results for a mean value
PRD of 0.53 (c.f. Table~\ref{TABLE1}) as this
value represents a good compromise between compression
performance and high visual
similitude of the recovered signal and the raw data.
Indeed, in \cite{EMW17} the quality of the
recovered signals giving rise to a mean value
PRD of 0.53 is illustrated in relation to the
high performance of automatic QRS complex detection.
However, the compression ratio of their  
method is low.
For the same mean value of PRD our CR is 5 times 
larger: 4.5 \cite{EMW17} vs 23.17 (Table~\ref{TABLE1}).
  As observed in Table~\ref{TABLE1} 
 the mean value of the local quantity
    $\prd$ is equivalent to the global value (PRD).
    Nevertheless the $\prd$ may differ for
    some of the segments in a record.
Fig.~\ref{prds} plots the $\prd$ for record 101 partitioned
into $Q=325$ segments of length $L=2000$ sample points.
Notice that there are a few
 segments corresponding to significantly larger values of
$\prd$ than the  others.
Accordingly,
with the aim of demonstrating the visual quality of
the recovered signals,
for each signal in the database we detect the
 segment  $q^\star$ of maximum distortion with respect to
the  $\prd$ as
\be
q^\star= \operatorname*{arg\,max}_{\substack{q=1,\ldots,Q}}\, \prd(q).
\ee
The left graphs of Fig.~\ref{Rec1} and Fig.~\ref{Rec2}
correspond to the segments of
maximum $\prd$ with respect to all the records
in the database and segments of length $L=2000$.
These are: the segment $25$ of records 101, when
applying the approximation approach a),
and segment $175$ of record 213
for approach  b). The upper waveforms  in
all the graphs are the raw ECG data. The lower waveforms
 are the corresponding approximations which have been shifted
down for visual convenience. The bottom lines in all the
graphs represent the absolute value of the difference
between the raw data and their corresponding approximation.
The right graphs of Fig.~\ref{Rec1} and \ref{Rec2}
have the same description as the top ones but the segments
 correspond to values of $\prd$ close 
to the mean value $\prd$ for the corresponding record.

It is worth commenting that the difference in
the results between approaches a) and b) is consequence of
the fact that the concomitant parameters are
set to approximate the whole database at a fixed
mean value PRD. In that sense, approach a) provides us with
 some flexibility (there are two parameters
to be fixed to match the required PRD) 
whereas for approach b) the only parameter 
($\Delta$) 
is completely determined by the required  
PRD. As observed in Table~\ref{PRD0},  when       
setting the parameter $\PRDO$ much smaller than the 
target PRD the approximation is only influenced 
 by the quantization parameter $\Delta$  and  
methods a) and b) coincide. Contrarily, when 
setting the $\PRDO$ too close to the 
target PRD the quantization parameter needs to be 
 significantly reduced, which affects the 
compression results. For a target $\PRD\ge 0.4$
 we  recommend to 
set $\PRDO$ as 70\%- 80\% of the required 
$\PRD$.  

For values of PRD$<0.4$ the storage approach is 
not as effective as for larger values of PRD. 
 This is noticeable in both Table~\ref{sphit1} and 
Table~\ref{TABLE6}. Another feature that 
appears for PRD$<0.4$ is that applying the  
entropy coding step, before saving the data in 
compressed HDF5 format, 
improves the CR much more than for larger 
values of PRD. This is because for PRD$<0.4$ the 
approximation fits noise and small details, 
 for which components in 
higher wavelet bands are required. 
Contrarily, for larger values of PRD the adopted 
uniform quantization keeps wavelet coefficients 
 in the first bands. As a result, through 
the proposed technique the location of 
the nonzero wavelet coefficients is encoded in 
 an array which contains mainly a long stream 
 of ones. For small values of PRD the array's length  
  increases to include different numbers. 
This is why the addition of an entropy coding step, 
such as Huffman coding which assigns smaller 
bits to the most frequent symbols, becomes more important. 
In any case, if the outputs are saved in HDF5 format, 
  adding the Huffman coding step is 
 beneficial. Nonetheless, since when implemented in software 
 the improvement comes at expense of computational time, 
 for PRD$>0.4$ this step can be avoided and the CR is 
 still very high. 

Comparisons with the conventional RL algorithm, in Table~\ref{TABLE6}, enhances the suitability of the proposal for 
storing the location of nonzero coefficients. 
A similar storage strategy has been successfully used 
with other approximation techniques for 
 compression of melodic music \cite{RNS17} and 
X-Ray medical images \cite{LRN18}.  
In this case the strategy is even more efficient,
because the approximation is realized using a basis and 
 on the whole signal, which intensifies the efficiency 
of the storage approach.
\section*{Conclusions}
\label{Conclu}
An effective and efficient method for 
compressing ECG signals 
has been proposed. The proposal was tested on the 
MIT-BIH Arrhythmia database, which gave rise to 
benchmarks improving upon recently reported results. 
The main feature of the method is its  
simplicity and  the fact that for values 
of PRD$>0.4$ a dedicated entropy coding 
to save the outputs can be avoided by 
saving the outputs of the algorithm in compressed
HDF5. This solution involves a time delay
which is practically negligible in relation to the
signal length: 0.14 s for compressing a 30 min record.
Two approaches for reducing wavelet coefficients have 
been considered. Approach b) arises from switching off 
in approach a) the selection of the largest wavelet 
coefficients before quantization. It was shown that, 
when approximating a whole database to obtain a 
fixed mean value of 
 PRD, approach a) may render a higher mean vale of CR 
when the target PRD is  greater the 0.4.

The role of the proposed Organization and Store strategy 
was highlighted by comparison  
with the conventional Run Length algorithm. Whilst the 
latter produces smaller CRs, the results are still good 
in comparison with previously reported benchmarks. 
This outcome leads to conclude that, using the 
 a wavelet transform on the whole signal,  
   uniform quantization for all the wavelet bands 
  works well in the design of 
a codec for lossy compression of ECG signals.\\

\noindent

{\bf{Note:}} The MATLAB codes for implementing the  whole
 approach have been made available on a dedicated website 
\cite{webpage}.


\subsection*{Acknowledgments} Thanks are due to
 K. Skretting, for making available the
 Huff06  MATLAB function \cite{Karl}, 
 which has been used for  
entropy coding, and  to P. Getreuer
 for the waveletcdf97 MATLAB 
  function \cite{PGcdf97} which has being used for 
  implementation of the CDF 9/7 wavelet transform.

\subsection*{Competing  interests}

The author declares no competing interests.

\subsection*{Author contribution statement}

The author is the only contributor to the paper.

\subsection*{Funding statement:}
There is no funding associated to this work.

\subsection*{Ethics statement:}
There is no Ethics issue related to this work.

\subsection*{Data availability:} 
The data used in this paper are available on\\
{\tt{https://physionet.org/physiobank/database/mitdb/}}\\
We have also placed the data, together with the 
software for implementing the proposed approach, 
on\\ {\tt{http://www.nonlinear-approx.info/examples/node012.htm}}

\end{document}